\documentclass{article}

\usepackage{arxiv}

\usepackage{bm}
\usepackage[utf8]{inputenc} 
\usepackage[T1]{fontenc}    
\usepackage{hyperref}       
\usepackage{url}            
\usepackage{booktabs}       
\usepackage{amsfonts}       
\usepackage{nicefrac}       
\usepackage{microtype}      
\usepackage{lipsum}
\usepackage{graphicx}
\usepackage{amsmath}
\usepackage{color}
\usepackage{dsfont}

\title{Quantifying the information lost in optimal covariance matrix cleaning}

\author{
 Christian Bongiorno$^{1,\dagger}$ \textmd{and} Lamia Lamrani\\
Université Paris-Saclay\\CentraleSupélec\\
Laboratoire de Mathématiques et Informatique pour la Complexité et les Systèmes\\
91192 Gif-sur-Yvette, France\\
 1.~\texttt{christian.bongiorno@centralesupelec.fr}\\
  $\dagger$. Corresponding author.
}

\usepackage{amsthm}

\theoremstyle{definition}

\begin{document}

\maketitle

\begin{abstract}
Obtaining an accurate estimate of the underlying covariance matrix from finite sample size data is challenging due to sample size noise.  In recent years, sophisticated covariance-cleaning techniques based on random matrix theory have been proposed to address this issue.  Most of these methods aim to achieve an optimal covariance matrix estimator by minimizing the Frobenius norm distance as a measure of the discrepancy between the true covariance matrix and the estimator. However, this practice offers limited interpretability in terms of information theory. To better understand this relationship, we focus on the Kullback-Leibler divergence to quantify the information lost by the estimator. Our analysis centers on rotationally invariant estimators, which are state-of-art in random matrix theory, and we derive an analytical expression for their Kullback-Leibler divergence. Due to the intricate nature of the calculations, we use genetic programming regressors paired with human intuition.  Ultimately, using this approach, we formulate a conjecture validated through extensive simulations, showing that the Frobenius distance corresponds to a first-order expansion term of the Kullback-Leibler divergence, thus establishing a more defined link between the two measures.
\end{abstract}

\section{Introduction}
Dependence characterization in multivariate statistics heavily relies on covariance matrix estimation, which accordingly is a tool of great importance in many areas such as physics \cite{andricioaei2001calculation}, neuroscience \cite{meng2023dynamic,ibanez2023noise, varoquaux2010brain, varoquaux2013learning,rahim2019population,liegeois2020revisiting,varoquauxestimating}, and finance \cite{laloux1999noise}. However, the sample covariance is remarkably noisy when the number of features is of the same order as the number of data points \cite{ledoit2003honey}. This situation is a frequent issue in many applications.
For example,  data on neural brain connections contains a large number of connections (features) but has small sample sizes \cite{meng2023dynamic,honnorat2022covariance}. In finance, constructing a well-diversified portfolio requires many assets (features). In contrast,  rapid shifts in financial market dependencies can only be captured by short calibration windows \cite{michaud1989markowitz,ledoit2022power}. These are just a select few examples, with many other fields also confronting this issue.

Numerous techniques have been developed to improve the estimation of noisy covariance matrices. Eigenvalue clipping \cite{laloux1999noise} and linear shrinkage \cite{ledoit2003honey,touloumis2015nonparametric} laid the groundwork in this area. Building upon these early works, a major improvement was the introduction of nonlinear shrinkage (NLS), initially proposed in Ref.~\cite{ledoit2011eigenvectors}, and further expanded in Ref.~\cite{ledoit2012nonlinear,ledoit2017direct}.  These methods, which are extensively reviewed in  Ref.~\cite{bun2017cleaning} and Ref.~\cite{ledoit2022power}, primarily originate from or can be understood within the framework of random matrix theory. Machine learning techniques have also been explored for enhancing covariance matrix estimation. Techniques such as cross-validation offer numerical solutions to the NLS problem for enhancing the estimator performances \cite{ibanez2023noise,  browne2000cross,bartz2016cross}. Alongside, alternative approaches like hierarchical clustering \cite{bongiorno2021covariance} and boosting methods \cite{bongiorno2022reactive} have independently emerged as effective methods in this domain. Generally speaking, applying constraints stabilizes the estimated covariance matrix, reducing its noise \cite{bun2017cleaning}. Notable examples include methods based on factor models \cite{fama1993common,de2021factor,fan2008high} and Bayesian estimation \cite{daniels1999nonconjugate}. The latter uses priors to effectively model the population covariance matrix \cite{barnard2000modeling,bouriga2013estimation}.

All these methods aim to reduce the noise in estimated covariance matrices. Yet, questions often arise about what these methods actually optimize and, more crucially, what they ideally should optimize. While the latter question is problem-specific, a make-do with good enough methods often prevails. For example, to minimize the variance of a portfolio, computations on the inverse covariance matrix are necessary. Nonetheless, employing NLS \cite{ledoit2011eigenvectors} on the covariance matrix itself is deemed acceptable as it minimizes the square distance between the filtered and the true matrix, i.e., the Frobenious error on the covariance matrix. Expectantly, in a realistic context characterized by finite sample size and time variability, minimizing the Frobenious error is not equivalent to minimizing the variance of a portfolio \cite{bongiorno2023non}.  Despite this, working with the Frobenius error is sometimes the only way to obtain analytical results.

More generally speaking, the Frobenius error lacks a clear interpretation when approaching the problem from an information theory perspective. Instead, the Kullback-Leibler (KL) divergence quantifies, by definition, the information lost when approximating the true distribution with an inferred one \cite{contreras2014asymptotic}. Alternative information metrics include the Jensen-Shannon divergence, Renyi divergence, and Wasserstein distance, among others. Each of these metrics has its own unique characteristics and applications. For example, the Jensen-Shannon divergence \cite{MENENDEZ1997307} is a symmetrized and smoothed version of KL divergence, making it more appropriate in some contexts where symmetry is desired. The Renyi divergence is a generalization of the KL divergence \cite{van2014renyi} that can emphasize different aspects of the distributional differences by adjusting the order parameter. The Wasserstein distance is particularly useful in contexts involving geometric differences and has properties that can be interpreted through optimal transport \cite{amari2018information}. The use of the KL divergence as a performance metric for evaluating filtered covariance matrix estimators was initially investigated in \cite{tumminello2007shrinkage,tumminello2007kullback} and then in \cite{bigot2022low}. They highlighted that the expected KL divergence is independent of the population (true) eigenvectors when the data is generated using Gaussian multivariate noise \cite{tumminello2007kullback}.  This unique attribute, alongside its capacity to measure information loss, earmarks the KL divergence as a valuable tool, especially in complex scenarios where eigenvectors are non-trivial. Despite its benefits, the application of KL divergence is limited in existing literature, probably because of its analytical challenges.
It is worth mentioning that the method in Ref.~\cite{ledoit2022quadratic} derives a covariance cleaning estimator by minimizing the inverse Stein loss, which is related to information-theoretic measures. Despite allowing for a more straightforward analytical estimator compared to non-linear shrinkage methods derived from random matrix theory \cite{ledoit2017numerical}, it achieves comparable performance, as confirmed by Monte Carlo simulations \cite{ledoit2022quadratic}.

We achieved further progress in applying information theory to covariance matrix cleaning. Specifically,  we conjectured the analytical expression for the KL divergence of optimal estimators. Given the ambiguity surrounding the ``optimal'' estimator, our investigation narrows its focus to the Rotationally Invariant Estimators (RIE) class. These estimators modify the eigenvalues of sample covariance matrices while maintaining the sample eigenvectors. In this simple setting, an optimal  RIE is known as the Oracle estimator if it minimizes the Frobenius error. Interestingly,  the Oracle estimator also minimizes the KL divergence for infinite sample sizes \cite{bongiornoUnpKL} as both cost functions have the same minimum. However, with a realistic finite sample size, opting to minimize the Frobenius norm over the KL divergence results in notable discrepancies.

To simplify our estimation problem, which involves computing the KL for the Oracle estimator, we focus on data distributed according to the inverse Wishart distribution. This serves as a starting point, and while it offers some analytical accessibility, it is by no means a limitation. The inverse Wishart distribution is one of the few for which analytic results are relatively easy to obtain. For example,  the optimal RIE estimator is the linear shrinkage 
\cite{Potters2020AFC}. Despite its simplicity, this distribution can approximate the eigenvalue distributions observed in real-world applications \cite{Potters2020AFC,turner2019incorporating}. Nevertheless, deriving the analytical expression of the KL divergence remains a challenge. As a shortcut, we employ Genetic Programming Regressors (GPR), a machine learning tool adept at solving symbolic regression problems \cite{koza1994genetic,koza2007genetic}. Operating through a tree representation of equations, it uses crossover and mutation operators to evolve potential solution populations based on Darwinian selection principles. After analyzing the GPR results, we employed human intuition to identify an incomplete series expansion within its output. Recognizing this allowed us to extend this partial sum to an infinite one, successfully obtaining the complete formula for the KL divergence of the optimal RIE in the inverse Wishart case.

The paper is organized as follows: in Section \ref{Section 2}, we set the definitions; in Section \ref{Section 3}, we derive analytically the KL divergence for sample covariance matrices; in Section \ref{Section 4},  we apply GPR to compute the KL divergence to the more complex case of the filtered covariance matrix, and finally, in Section \ref{Section 5}, we link the KL divergence to the Frobenius error.

\section{Definitions}
\label{Section 2}
\subsection{Kullback-Leibler Divergence}

The KL divergence, also known as relative entropy, measures how one probability distribution diverges from a second, expected probability distribution. Given probability measures $P$ and $Q$ over a set $\mathcal{X}$, the KL divergence is formally defined as \cite{kullback1951information,cover1991information}:
\begin{equation}\label{eq:KL}
\textrm{KL}(P||Q) := \int_{\mathcal{X}} \left( \log \frac{P}{Q} \right) dP = \mathbb{E}\left[ \log \left( \frac{P(x;{\bf C})}{Q(x;{\bf S})}\right) \right]_{P(x;{\bf C})}.
\end{equation}

Conceptually, the KL divergence quantifies the discrepancy between the two probability measures $P$ and $Q$. It is a non-negative quantity; however, it is important to note that the KL divergence is not a distance metric as it is not symmetric, i.e., $\textrm{KL}(P||Q) \neq \textrm{KL}(Q||P)$ in general.

The KL divergence represents the average logarithmic difference between the probabilities $P$ and $Q$, where the average is taken using the probabilities $P$. Consequently, it provides an estimate of the expected amount of information (in bits, for logarithm base 2, or nats, for natural logarithm) lost when $Q$ is used to approximate $P$.

In the case of multivariate normal variables with true population covariance $\bm{C}$ and  estimator $\bm{S}$, the KL divergence simplifies to \cite{tumminello2007kullback}
\begin{equation}\label{eq:KLcomp}
\textrm{KL}(\bm{C}||\bm{S}) = \frac{1}{2}\left(\textrm{Tr}\left(\bm{S}^{-1} \bm{C}\right) + \log\left(\frac{\det(\bm{S})}{\det(\bm{C})}\right) - n\right),
\end{equation}
where $Tr(\bullet)$ is the trace operator.

This work focuses on a normalized version of the KL divergence. By dividing the KL divergence by the dimension of the covariance matrix $n$, we obtain a scaled measure of the divergence, which is particularly useful for high-dimensional problems:
\begin{equation}\label{eq:KLnorm}
\overline{\textrm{KL}}(\bm{C}||\bm{S}) = \frac{\textrm{KL}(\bm{C}||\bm{S})}{n}.
\end{equation}

This normalized KL divergence provides a more interpretable measure of the information loss per dimension when the estimator $\bm{S}$ is used to approximate the true population covariance $\bm{C}$.

\subsection{Populations, Samples, and Oracle Estimator}

Let $n\in\mathbb{N}^{*}$ and $p>0$. Let us define $q^{*}=\frac{p}{1+p}$ and $t^*=\lfloor \frac{n}{q^{*}}\rfloor$. Let us consider $(\bm{m}_{i})_{1\leq i\leq t^{*}}$ iid vectors of $\mathbb{R}^{n}$ such that $\bm{m}_{1}\sim\mathcal{N}(0,\bm{\Sigma})$. 
Since we focus on the white inverse Wishart case, $\bm{\Sigma}=\bm{\mathds{1}}$.
We then define $\bm{M}=(\bm{m}_{1},...,\bm{m}_{t^*})\in\mathbb{R}^{n\times t^*}$. 
Then, 
\begin{equation}    
\bm{W}_{q^{*}}=\frac{1}{t^*}\bm{MM}^{\top}
\end{equation}
is a Wishart $(n,q^{*})$ matrix and 
\begin{equation}
\bm{\mathcal{W}}^{-1}_{np}=(1-q^{*})\bm{W}^{-1}
\end{equation}
is an Inverse Wishart with parameters $(n,p)$ \cite{Potters2020AFC}. 
The constant of $(1-q^{*})$ is added so that the expected value of the normalized trace of an Inverse Wishart is equal to $1$. 

The Inverse Wishart distribution displays a spread-out distribution of the eigenvalues when $p>0$ and becomes marked when $p$ is very large (and therefore $q^*$ approaches one); such behavior cannot be achieved with a simpler Wishart matrix. In addition,  the eigenvectors of a white Wishart matrix are distributed according to the Haar measure on the orthogonal group \cite{bouferroum2013eigenvectors}.
    
The associated data matrix $\bm{X}$, of dimensions $n \times t$, is generated from a centered multivariate normal distribution with population covariance matrix $\bm{C}:=\bm{\mathcal{W}}^{-1}_{np}$. Following this, after removing the average of each column of $\bm{X}$, the sample covariance matrix is defined as
\begin{equation}
\bm{E} = \frac{1}{t} \bm{X X}^{\top},
\end{equation}
with $q:=\frac{n}{t}$ the ratio between the size of the matrix and the number of observations.

This sample covariance matrix $\bm{E}$ is decomposed spectrally \cite{johnson2002applied} as
\begin{equation}
\bm{E} = \bm{V \Lambda V}^{\top},
\end{equation}
where $\bm{V}$ is the matrix of eigenvectors and $\bm{\Lambda}$ is the diagonal matrix of eigenvalues. From this spectral decomposition, we can extract the Oracle eigenvalues as
\begin{equation}
\bm{\Lambda}_O = \textrm{diag}\left( \bm{V}^\top \bm{C V} \right),
\end{equation}
where the operator $\textrm{diag}(\bullet)$ nullifies the off-diagonal elements of the matrix \cite{Potters2020AFC}.

With the Oracle eigenvalues $\bm{\Lambda}_O$ in hand, we define the Oracle estimator $\bm{\Xi}(\bm{\Lambda}_O | \bm{E})$ as
\begin{equation}
\bm{\Xi}(\bm{\Lambda}_O | \bm{E}) := \bm{V \Lambda}_O \bm{V}^{\top}.
\end{equation}

This Oracle estimator represents an idealized estimator that is aware of the true population covariance. $\bm{\Xi}(\bm{\Lambda}_O | \bm{E})$ is the Rotational Invariant Estimator (RIE) that minimizes the Frobenius norm distance with the population matrix $\bm{C}$ \cite{ledoit2011eigenvectors,Potters2020AFC,bongiornoUnpKL}. In what follows, we simplify the notation by omitting the dependence on $\bm{E}$ and refer to the estimator simply as $\bm{\Xi}(\bm{\Lambda}_O )$.  

In the case of multivariate normal or infinite multivariate t-students variables, it has been proved \cite{bongiornoUnpKL} that $\bm{\Xi}(\bm{\Lambda}_O)$ is also the RIE that minimizes the KL divergence $\textrm{KL}(\bm{C}||\bm{\Xi}(\bm{\Lambda}))$. This finding implies that the Oracle estimator, derived from the inverse Wishart population covariance matrix is, in some idealized context, optimal for minimizing both the Frobenius norm and the KL divergence. Therefore, it is the proper tool for covariance matrix estimation in high-dimensional settings.

\section{Quantifying Information Loss for Sample Covariances}
\label{Section 3}

In this section, we quantify the expected loss of information analytically when the population matrix $\bm{C}$ is approximated by the sample estimator $\bm{E}$.

For the sake of generality, we consider $\bm{C} \in \mathbb{R}^{n \times n}$ a generic positive defined population covariance matrix, i.e., not necessarily a white inverse Wishart, and $\bm{E}$, the sample covariance, obtained from $\bm{X} \in \mathbb{R}^{n \times t}$  with $t>n$ multinomial variables with population covariance matrix $\bf C$.

In the limit of large $n$, this expected information loss, as measured by the normalized KL divergence, is given by
\begin{equation}
\label{8}
\mathbb{E}[ \overline{\textrm{KL}}(\bm{C}||\bm{E})] = \frac{1-q}{2 q} \log \left(\frac{1}{1-q} \right) + \frac{1}{2(1-q)} -1.
\end{equation}

Let us briefly explain where this equation comes from. Firstly, the normalized KL is defined as  \cite{tumminello2007kullback}
\begin{equation}
\mathbb{E}[ \overline{\textrm{KL}}(\bm{C}||\bm{E})]=\frac{1}{2}\mathbb{E}\left[\tau(\bm{E}^{-1}\bm{C})-1+\frac{1}{n}\log\left(\frac{\det(\bm{E})}{\det(\bm{C})}\right)\right]
\end{equation}
where $\tau(\bullet)=\frac{1}{n}Tr(\bullet)$ is the normalized trace.

By noticing that $\bm{E}=\sqrt{\bm{C}} \bm{W}_{q}\sqrt{\bm{C}}$, with $\bm{W}_{q}$ a white Wishart generated by $t=\lfloor \frac{n}{q}\rfloor$ observations and independent of $\bm{C}$, our expression simplifies to

\begin{equation}
\mathbb{E}[ \overline{\textrm{KL}}(\bm{C}||\bm{E})]=\frac{1}{2}\mathbb{E}\left[\tau(\bm{W}_{q}^{-1})-1+\frac{1}{n}\log(\det(\bm{W}_{q}))\right].
\end{equation}

We can already notice that the KL does not depend on the choice of the population matrix when the observations are generated by a Gaussian multiplicative distribution.

The first term is known \cite{Potters2020AFC}:
\begin{equation}
\mathbb{E}[\tau(\bm{W}_{q}^{-1})]=\frac{1}{1-q}.
\end{equation}

To compute the second term, we can use the fact that the spectral density of a white Wishart follows a Mar\v cenko-Pastur distribution in the asymptotic limit ($n,t\rightarrow\infty$ with $\frac{n}{t}\rightarrow q$) \cite{plerou2002random} and were able to compute it using Mathematica \cite{Mathematica}, which yields

\begin{equation}
\mathbb{E}\left[\frac{1}{n}\log(\det(\bm{W}_{q}))\right]=\frac{1-q}{q}\log\left(\frac{1}{1-q}\right)-1.
\end{equation}

The above equation provides a measure of how much information, on average, is lost when the population covariance matrix $\bm{C}$ is approximated by the sample covariance matrix $\bm{E}$.

Furthermore, suppose we are interested in comparing the loss of information between two matrices, $\bm{E}_{\textrm{in}}$ and $\bm{E}_{\textrm{out}}$, both derived from the same population covariance $\bm{C}$ but with different parameters $q_{\textrm{in}}$ and $q_{\textrm{out}}$ and independent data points. The corresponding loss of information is given by
\begin{align}
\label{kl_in_out}
\mathbb{E}[\overline{\textrm{KL}}(\bm{E}_{\textrm{out}}||\bm{E}_{\textrm{in}})] &= \frac{1-q_{\textrm{in}}}{2 q_{\textrm{in}}} \log \left(\frac{1}{1-q_{\textrm{in}}} \right) 
- \frac{1-q_{\textrm{out}}}{2 q_{\textrm{out}}} \log \left(\frac{1}{1-q_{\textrm{out}} } \right)  \nonumber\\&+\frac{1}{2 (1-q_{\textrm{in}}) } -\frac{1}{2}. 
\end{align}

This equation provides a means of assessing the relative loss of information when using different sample covariance matrices to approximate the same population covariance matrix.

It is a consequence of Eq.~\eqref{8}, let us briefly explain why

\begin{align}
\mathbb{E}[ \overline{\textrm{KL}}(\bm{E}_{\textrm{out}}||\bm{E}_{\textrm{in}})] & =\frac{1}{2} \mathbb{E}\left[\tau(\bm{E}^{-1}_{\textrm{in}}\bm{E}_{\textrm{out}} )-1+\frac{1}{n}\log\left(\frac{\det(\bm{E}_{\textrm{out}})}{\det(\bm{E}_{\textrm{in}})}\right)\right] \nonumber \\
& =\frac{1}{2} \mathbb{E}\left[\tau(   \bm{W}^{-1}_{\textrm{in}}\bm{W}_{\textrm{out}})-1+\frac{1}{n} \log\left(\det(\bm{W}_{\textrm{in}}\right)-\frac{1}{n}\log(\det(\bm{W}_{\textrm{out}}))\right].
\end{align}

Using the fact that the white Wisharts $\bm{W}_{\textrm{in}}$ and $\bm{W}_{\textrm{out}}$ are asymptotically free in the large limit and the moments of the log Marcenko-Pastur distribution, we obtain

\begin{align}
\mathbb{E}[ \overline{\textrm{KL}}(\bm{E}_{\textrm{out}}||\bm{E}_{\textrm{in}})]&=\frac{1}{2} \left(\frac{1}{1-q_{\textrm{in}}}-1+\frac{1-q_{\textrm{in}}}{q_{\textrm{in}}} \log \left(\frac{1}{1-q_{\textrm{in}}} \right)-1\right)\nonumber  \\ & + \frac{1}{2}\left(- \frac{1-q_{\textrm{out}}}{q_{\textrm{out}}} \log \left(\frac{1}{1-q_{\textrm{out}}} \right) +1\right) \nonumber
\\
&=\frac{1}{2(1-q_{\textrm{in}})}+\frac{1-q_{\textrm{in}}}{2 q_{\textrm{in}}} \log \left(\frac{1}{1-q_{\textrm{in}}} \right) \nonumber\\ &- \frac{1-q_{\textrm{out}}}{2q_{\textrm{out}}} \log \left(\frac{1}{1-q_{\textrm{out}}} \right) -\frac{1}{2}. 
\end{align}

From the law of large numbers,  when $q_{\textrm{out}}\rightarrow 0$,  $\bm{E}_{\textrm{out}}$ converges almost surely to $\bm{C}$. Therefore, 
\begin{equation}
    \lim_{q_{\textrm{out}} \to 0} \mathbb{E}[ \overline{\textrm{KL}}(\bm{E}_{\textrm{out}}||\bm{E}_{\textrm{in}})] = \mathbb{E}[ \overline{\textrm{KL}}(\bm{C}||\bm{E}_{\textrm{in}})].
\end{equation}
This can be easily proved by Taylor's expansion
\begin{equation}
    \lim_{q_{\textrm{out}} \to 0} -\frac{1-q_{\textrm{out}}}{2q_{\textrm{out}}}\log \left( \frac{1}{1-q_{\textrm{out}}}\right)=-\frac{1}{2}.
\end{equation}

However, it is worth stressing that $q_{in}\rightarrow q_{\textrm{out}}$ does not imply that $\bm{E}_{\textrm{out}}$ and $\bm{E}_{\textrm{in}}$ become equal; we only know that they become sample covariances generated by the same number of observations with independent data i.e. that both sample covariances have the same level of noise with respect to the population covariance matrix. Therefore eq.\ref{kl_in_out} does not vanish when $q_{\textrm{in}}=q_{\textrm{out}}$.

While the expected KL divergence in the finite sample size regime is also known \cite{tumminello2007kullback}, it involves sums over digamma functions, which are not conducive for analytical computations.

It is worth noting that these results are applicable to any population covariance matrix and, thus, not restricted to the Wishart distribution. This broad applicability renders our findings useful in a wide range of practical scenarios beyond the scope of this work.

\section{Kullback-Leibler Divergence for the Oracle of an Inverse Wishart}\label{sec:procedure}
\label{Section 4}
\subsection{Mathematical Challenges for Analytical Derivation}

The process of optimally deriving the KL divergence of a RIE that converges to the Oracle for an inverse Wishart matrix presents a variety of mathematical challenges \cite{Potters2020AFC}.

For the white inverse Wishart prior, it is known that the optimal RIE corresponds to a linear shrinkage of the sample covariance \cite{Potters2020AFC} and is given by 
\begin{equation}
\bm{\Xi}( \bm{\Lambda}_O | \bm{E}) = r(\bm{E} - \bm{\mathds{1}}) +\bm{\mathds{1}},
\end{equation}
where
\begin{equation}\label{eq:r}
r = \frac{n p}{n ( p+q )-p q}.
\end{equation}

Therefore, the KL divergence related to this estimator is given by

\begin{equation}\label{eq:KLoracle}
    \textrm{KL}(\bm{C}||\bm{\Xi}(\bm{\Lambda}_O)) = \frac{1}{2}\left(\textrm{Tr}\left(\bm{\Xi}^{-1}(\bm{\Lambda}_O) C\right) + \sum_{i=1}^n \left( \log \Lambda^O_i -\log \zeta _i \right) - n\right).
\end{equation}

Here, $\zeta_i$ denotes the eigenvalues of the population matrix.
The determination of the expected value in the above equation requires multiple approximations. For instance, obtaining the expected value of the logarithm necessitates a Taylor expansion due to the nonlinearity of the logarithmic function. Furthermore, the calculation of the trace correlation needs intricate knowledge of the interdependence among the eigenvalues of the sample estimator.

These mathematical complexities underline the inherent challenge in obtaining an analytical solution for the KL divergence in the case of the Oracle of a white inverse Wishart. In essence, the interplay of nonlinearities, random variables, and dependencies presents considerable obstacles to the derivation of an analytical solution. Therefore, we seek a novel approach to tackle this problem, which we explore in the following sections.

\subsection{Symbolic Machine Learning}
In order to derive the expected value of the normalized KL divergence for an arbitrary white inverse Wishart matrix $\bm{\mathcal{W}}^{-1}_{np}$ in the large $n$ regime, we tried genetic symbolic regressor approach. Symbolic regression is a type of regression analysis that searches for mathematical expressions that best fit a given dataset without specifying the form of the model in advance. A Genetic Programming (GP) symbolic regressor \cite{stephens2016introduction} is therefore a genetic algorithm aiming to derive analytical solutions by evolving mathematical expressions in a way analogous to natural evolution. The working logic of the GP-based symbolic regressor is as follows: first, the algorithm starts with a randomly generated population of mathematical expressions, also called individuals or candidate solutions. Each expression is represented as a tree structure, where nodes correspond to mathematical operators (e.g., addition, subtraction, multiplication, division), and leaves represent variables (in our case, the parameters $q$ and $r$) or constants. Then, each candidate expression is evaluated using a fitness function. In our application, the fitness function was the mean squared error between the expression's output and the normalized KL divergence obtained numerically.  Candidate expressions are selected for reproduction based on their fitness. Different selection strategies might be used; in our work, we adopted a tournament selection. In a tournament selection, a subset of expressions is chosen randomly, and the best among them is selected for reproduction. Pairs of selected expressions are combined via the crossover operator by swapping subtrees between them, creating new offspring expressions. Random changes are also introduced to individual expressions via the mutation operator by altering nodes or subtrees, usually at a rate very low rate. While the crossover allows the algorithm to explore new areas of the solution space by recombining existing building blocks, the mutation adds diversity to the population and helps prevent premature convergence on suboptimal solutions. In order to overcome overcomplicated solutions, a parsimony coefficient is generally introduced. This coefficient is used in the fitness function to subtract a small value proportional to the size of the expression. This process is iterated for a certain number of steps, also known as generations, and the best individual is finally considered the output of the algorithm. 

We initialize our dataset by sampling $5,000$ Inverse-Wishart matrices $\bm{C}$ with randomly chosen parameters $q^{*} \in (0,1)$ and $n=1,000$, and subsequently extracting sample covariances $\bm{E}$ from these matrices with random $q \in (0,1)$. For each pair of matrices, we computed the Oracle estimator $\bm{\Xi}(\bm{\Lambda}_O)$ and the related KL according to Eq.~\eqref{eq:KLoracle}. To improve stability, for every pair of chosen $(q,q^*)$, we repeated $500$ times the sampling of $\bm{C}$ and $\bm{E}$, and we averaged the resulting KL. 

We initiated the process with a population of $50,000$ potential solutions, set the parsimony coefficient to $10^{-4}$, and allowed for $40$ generation rounds. All other parameters were left at their default settings \cite{stephens2016introduction}. A preliminary analysis of the simulation data suggested that we should use the parameters $q$ and $r$ (as defined in Eq.~\eqref{eq:r}) as input for the regressor, with the goal of minimizing the mean square error of the simulated normalized KL divergence obtained from Eqs. \eqref{eq:KLcomp} and \eqref{eq:KLnorm}. The codes to generate the data and calibrate the GPR are available at GitHub \cite{repository}.

Because of the tendency of genetic algorithms to become entrapped in local minima if not adequately tuned, we ran a total of four independent rounds. From these, we identified a generic term for the normalized KL divergence
\begin{equation}\label{eq:secondapproxr}
    \mathbb{E}[\overline{\textrm{KL}}(\bm{\mathcal{W}}^{-1}_{np}|| \bm{\Xi}(\bm{\Lambda}_O) )] \approx \frac{1}{4} r q - \left( \frac{1}{4}\ r q\right)^2.
\end{equation}
This equation bears a resemblance to a second-order Taylor expansion. Considering the large $n$ limit of
\begin{equation}
\lim_{n \to \infty} r q = \frac{q^{*} q}{q^{*} + q - q^{*} q}.
\end{equation}
Eq.~\eqref{eq:secondapproxr} can be transformed into
\begin{equation}\label{eq:secondapprox}
    \mathbb{E}[\overline{\textrm{KL}}(\bm{\mathcal{W}}^{-1}_{np}|| \bm{\Xi}(\bm{\Lambda}_O) )] \approx \frac{1}{4}\frac{q^{*} q}{q^{*} + q - q^{*} q} - \left( \frac{1}{4}\frac{q^{*} q}{q^{*} + q - q^{*} q} \right)^2.
\end{equation}

To validate our assumption, we performed a second round of tests, this time expanding the range of $q^{*} \in (0,1)$ and $q \in (0,7)$. It's important to note that when $q>1$, the sample matrix $E$ will be singular, but not the Oracle $\bm{\Xi}(\bm{\Lambda}_O)$. Our analysis of this new data set revealed a discrepancy regarding the prediction of \eqref{eq:secondapprox}.

\begin{figure}
    \centering
    \includegraphics[width=0.6\columnwidth]{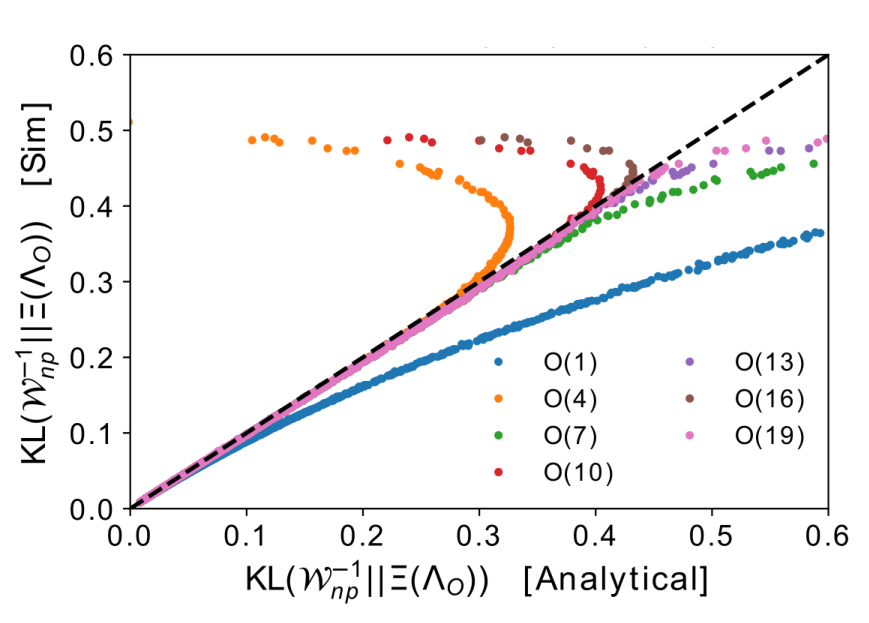}
    \caption{Comparison of the convergence of the series for different orders of approximation of Eq.~\eqref{eq:taylor} with parameters $n=1000$, $q^*\in(0,1)$ and $q\in (0,7)$.}
    \label{fig:seriesconv}
\end{figure}

We attempted to expand Eq.~\eqref{eq:secondapprox} into a suitable progression:
\begin{equation}\label{eq:taylor}
    \mathbb{E}[\overline{\textrm{KL}}(\bm{\mathcal{W}}^{-1}_{np}|| \bm{\Xi}(\bm{\Lambda}_O) )] = \sum_{k=1}^{\infty} (-1)^{k-1}\left( \frac{1}{4}\frac{q^{*} q}{q^{*} + q - q^{*} q} \right)^k,
\end{equation}
which yielded an exceptional match with the predictions as the approximation order increased, as depicted in Fig.~\ref{fig:seriesconv}.

Moreover, it is here possible to recognize the Taylor-expansion of the function $x\rightarrow \frac{1}{1+x}$ as
\begin{eqnarray}\label{eq:taylor simplified}
    \mathbb{E}[\overline{\textrm{KL}}(\bm{\mathcal{W}}^{-1}_{np}|| \bm{\Xi}(\bm{\Lambda}_O) )] =&& \frac{1}{4}rq \sum_{k=0}^{\infty} (-1)^{k}\left( \frac{1}{4}rq \right)^k
    \nonumber\\
    &&=\frac{1}{4}rq\frac{1}{1+\frac{1}{4}rq}.
\end{eqnarray}

Finally, by replacing $r$ by its value in the high dimension limit, we obtain
\begin{equation}\label{eq:kl assympt}
    \mathbb{E}[\overline{\textrm{KL}}(\bm{\mathcal{W}}^{-1}_{np}|| \bm{\Xi}(\bm{\Lambda}_O) )] =\frac{pq}{4p+4q+pq},
\end{equation}
which converges if and only if
\begin{equation}\label{eq:region}
 rq=\frac{q^{*} q}{q^{*} + q - q^{*} q} <  4.
\end{equation}

Let us remind that the shrinkage coefficient $r$ necessarily lies in $[0,1]$ interval; therefore, the divergence can only occur when $q>4$, which means that the number of features should be at least $4$ times larger than the number of observations. 

\begin{figure}[ht]
    \centering
    \includegraphics[width=6cm]{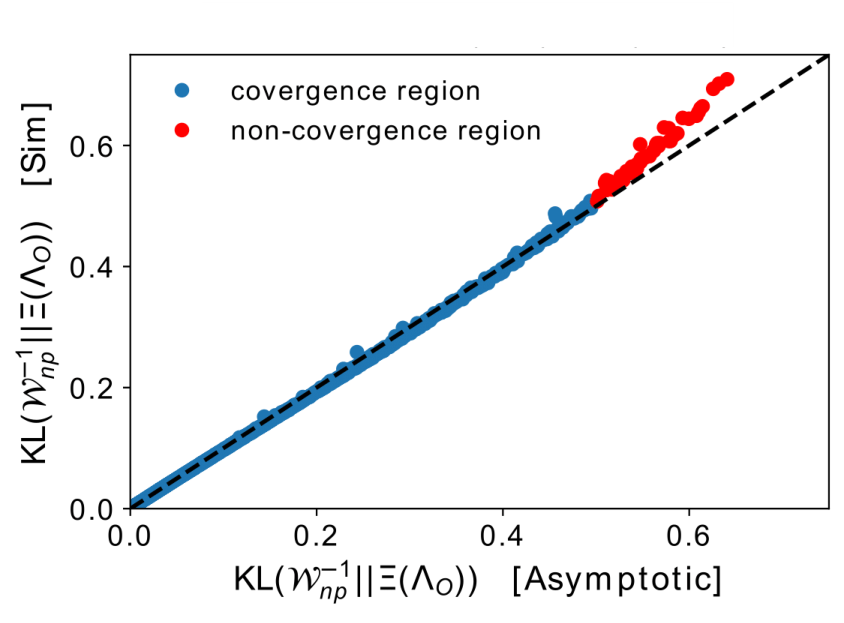}
    \includegraphics[width=6cm]{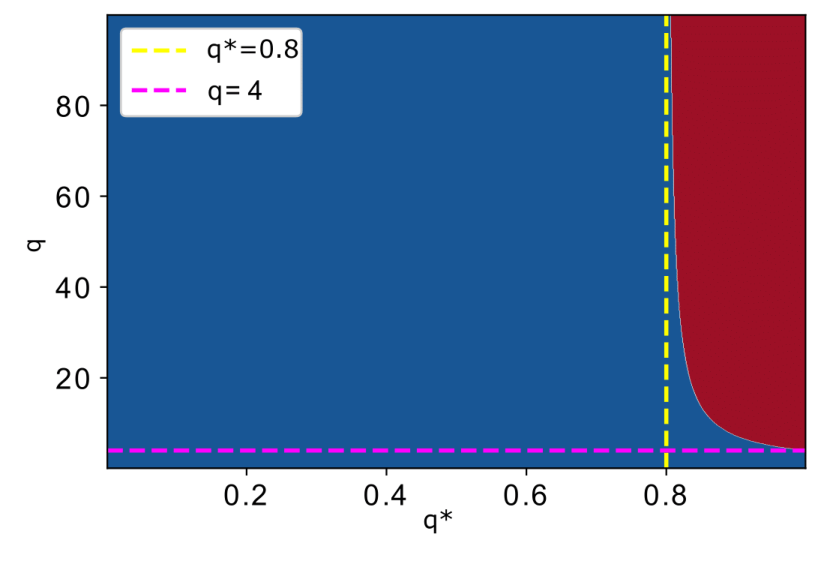}
    
    \caption{The left panel shows the agreement of the asymptotic equation \eqref{eq:secondapprox} for parameters $n=1000$, $q^{*}\in(0,1)$ and $q\in(0,7)$; the right panel shows the convergence region of \eqref{eq:region}; in the red region, the inequality is not respected.}
    \label{fig:asymtotic}
\end{figure}

The left panel of Fig.~\ref{fig:asymtotic} shows strong agreement with Eq.~\eqref{eq:secondapprox}. The right panel of Fig.~\ref{fig:asymtotic} showcases the convergence region defined by Eq.~\eqref{eq:region}, with two asymptotic values $q^{*}=0.8$ and $q=4$.

\section{Analysis of the Frobenius Error}
\label{Section 5}
In the previous sections, we have derived the KL divergence in the context of the optimal linear shrinkage estimators. However, a substantial part of the existing literature focuses on the Frobenius error. Both the KL divergence and Frobenius error could provide different perspectives and conclusions. 

Given our population matrix $\bm{C}$ and the established assumption that it follows a white inverse Wishart distribution with parameter $p$, we can write the Frobenius error of the Oracle as \cite{Potters2020AFC}
\begin{equation}
\label{31}
\overline{\textrm{Frob}}((\bm{C}|| \bm{\Xi}(\bm{\Lambda}_O)) =\tau((\bm{\Xi}(\bm{\Lambda}_{O})-\bm{C})^{2})
 \end{equation}
where $\tau$ is the normalized trace.

This follows directly from the definition of the Frobenius norm and the fact that the optimal linear shrinkage is equal to the Oracle when estimating a population matrix using the sample covariance $\bm{E}$. Also, it is known that the expected value of the Frobenius error of estimation of a white inverse Wishart is the following \cite{Potters2020AFC,llamrani}
\begin{equation}
\mathbb{E}[\overline{\textrm{Frob}}((\bm{C}|| \bm{\Xi}(\bm{\Lambda}_O))] = (1-r)^{2}p+ qr^{2}.
\end{equation}

Given that $r=\frac{p}{p+q}$ in the large limit, we finally obtain
\begin{equation}
\mathbb{E}[\overline{\textrm{Frob}}((\bm{C}|| \bm{\Xi}(\bm{\Lambda}_O))] = \frac{pq}{p+q}.
\end{equation}

Importantly, we note that if $p$ or $q$ are small,
\begin{equation}
\mathbb{E}[\overline{\textrm{KL}}(\bm{C}|| \bm{\Xi}(\bm{\Lambda}_O))] \sim \frac{1}{4}\mathbb{E}[\overline{\textrm{Frob}}((\bm{C}|| \bm{\Xi}(\bm{\Lambda}_O)))].
\end{equation}

This shows a clear relationship between the Frobenius error and the KL divergence in the case of a white inverse Wishart population matrix. From the Taylor-Expansion of the KL divergence (eq.~\eqref{eq:secondapproxr}), we can understand where the $\frac{1}{4}$ factor term comes from, as we notice that the first-order term of the expected KL, $\frac{1}{4}rq$, coincides exactly with $\frac{1}{4}$ of the expected Frobenius error

\begin{align}\label{eq:taylor frob}
    \mathbb{E}[\overline{\textrm{KL}}(\bm{\mathcal{W}}^{-1}_{np}|| \bm{\Xi}(\bm{\Lambda}_O) )]&=\sum_{k=1}^{\infty} (-1)^{k-1}\left( \frac{1}{4}rq \right)^k \nonumber
    \\&= \sum_{k=1}^{\infty} (-1)^{k-1}\left( \frac{1}{4}\mathbb{E}[\overline{\textrm{Frob}}(\bm{C}|| \bm{\Xi}(\bm{\Lambda}_O)]) \right)^k.
\end{align}

As a consequence, in the case of a white inverse Wishart population matrix, the expected Frobenius error of estimation is proportional to the first-order approximation of the KL divergence. We point out that while the Frobenius error is always finite, the KL divergence can be infinite at the same time.

\section{Conclusions}

In this work, we presented a novel method for computing the analytical expected value of the KL divergence between a population covariance matrix and its optimal estimator, specifically for an inverse Wishart matrix and Gaussian data in a high-dimensional setting.

Our findings show that the conjectured expected KL divergence can be derived from a convergent series expansion. The convergence of this expansion is condition-dependent, leading to new research questions. Additionally, by comparing the Frobenius error of estimation between the Oracle and the population matrix, we observed that the Frobenius error corresponds to a quarter of the first-order term of the KL divergence.

Instead of a strictly mathematical approach, often limited by technical challenges, we used a machine learning-based strategy employing symbolic regression through a genetic algorithm. This strategy simplified the derivation of an explicit expression for the KL divergence, bypassing the analytical computation challenges. The application of machine learning in this scenario underscores its utility in tackling complex mathematical problems, especially when conventional mathematical approaches fall short. 
However, the current approach is not yet standardized and its application to different cases should be carefully tailored to the specific characteristics of the data. Future work should focus on standardizing the approach and exploring its generalizability.

Future research could expand our methodology to various types of matrices or distributions, especially non-Gaussian ones, where MSE does not provide a reliable loss function for optimization.

\subsection*{Acknowledgments}
We want to thank Damien Challet for particularly useful discussions and advice which helped improve the paper. This work was performed using HPC resources from the ``Mésocentre'' computing center of CentraleSup\'elec and \'Ecole Normale Sup\'erieure Paris-Saclay supported by CNRS and R\'egion \^Ile-de-France (\url{http://mesocentre.centralesupelec.fr/}). 
\subsection*{Funding Disclosure}

This research did not receive any specific grant from funding agencies in the public, commercial, or not-for-profit sectors.

\bibliographystyle{unsrt}  
\bibliography{references}

\end{document}